\documentclass[aps,a4paper,amssymb,amsmath,reprint,showpacs,superscriptaddress]{revtex4-1}
\usepackage{bm}
\usepackage{hhline}
\usepackage{amsmath}
\usepackage{amssymb}
\usepackage{mathdots}
\usepackage{tabularx}
\usepackage{graphicx}
\usepackage{hyperref}
\usepackage{bbding}
\usepackage{tikz}
\usepackage{bbding}
\usepackage{braket}
\usepackage{bm}
\usepackage{physics}
\usepackage{mathdots}

\begin{document}
	\title{Unconventional Josephson effects in {\it PT}-symmetric antiferromagnetic bilayers}
	\author{Jin-Xin Hu}\thanks{jhuphy@ust.hk}
        \affiliation{Department of Physics, Hong Kong University of Science and Technology, Clear Water Bay, Hong Kong, China} 
        \author{Mengli Hu}
        \affiliation{Institute for Theoretical Solid State Physics, IFW Dresden, Helmholtzstrasse 20, 01069 Dresden, Germany}

 \author{Ying-Ming Xie}\thanks{yxieai@connect.ust.hk}
    \affiliation{RIKEN Center for Emergent Matter Science (CEMS), Wako, Saitama 351-0198, Japan}
 	\author{K. T. Law}\thanks{phlaw@ust.hk}
\affiliation{Department of Physics, Hong Kong University of Science and Technology, Clear Water Bay, Hong Kong, China} 	
	\date{\today}
	\begin{abstract}

We propose that unconventional Josephson effects can typically emerge in {\it PT}-symmetric antiferromagnetic (AFM) bilayer systems. When proximitized by a conventional superconductor, these heterostructures host dominant interlayer Cooper pairing that features a distinctive spin texture enabled by the strong exchange field. Specifically, we demonstrate a novel mechanism for electrically tunable 0-$\pi$ oscillations in lateral Josephson junctions, controlled by an out-of-plane electric displacement field. This behavior originates from field-induced finite-momentum Cooper pairing, a hallmark of the unique layer-pseudospin structure in {\it PT}-symmetric AFM bilayers. Furthermore, we introduce a Josephson giant magnetoresistor based on these exotic spin-layer-locked Cooper pairs, in which the supercurrent exhibits a strong dependence on the internal N\'{e}el order. Our findings establish {\it PT}-symmetric AFM bilayers as a versatile platform for phase-controllable Josephson junctions and superconducting magnetic random-access memory, with promising applications in superconducting circuits and ultralow-power computing.

	\end{abstract}
	\maketitle

\section{Introduction}
The quest for fault-tolerant quantum computing has spurred intense research into topological superconductors~\cite{fu2008superconducting,nayak2008non,qi2010chiral,alicea2012new} and phase-controlled Josephson junctions (JJs)~\cite{ioffe1999environmentally,blatter2001design,you2002scalable,yamashita2005superconducting,buzdin2005peculiar,jorgensen2007critical,gingrich2016controllable}. In conventional JJs, it is known that the supercurrent $I_s$ follows $I_s=I_c\sin(\phi)$, where $\phi$ is the phase difference between superconducting electrodes and $I_c$ is the critical current~\cite{josephson1962possible,josephson1964coupled}. Introducing magnetism into the junction barrier can induce a $\pi$-phase shift, creating a ground-state current-phase relation $I_s=I_c\sin(\phi+\pi)$---a hallmark of $\pi$-junctions~\cite{robinson2006critical,linder2008supercurrent}. Such junctions are the building blocks of topologically protected superconducting qubits~\cite{gladchenko2009superconducting,kitaev2006protected,gyenis2021experimental,paolo2019control,yamashita2005superconducting,brooks2013protected,groszkowski2018coherence,smith2020superconducting,guo20220,rajpoot2022tunable}.

Conventional paradigms for the 0-$\pi$ transitions in ferromagnetic JJs have been extensively studied, which are governed by temperature and the thickness of the ferromagnetic layer~\cite{ryazanov2001coupling,kontos2002josephson,born2006multiple,barash2002interplay,robinson2007zero,frolov2006josephson,weides2006high,weides20060,Bergeret2005}. However, the practical application of ferromagnetic JJs faces outstanding technical challenges, such as preventing stray-field–induced screening supercurrents and Abrikosov vortex nucleation in adjacent superconductors. This has evoked recent interest in exploring long-range supercurrents through antiferromagnets within lateral JJs~\cite{Jeon2021}. Due to the essentially zero net magnetization, they produce no stray fields. Recent theoretical studies also suggest that altermagnets could provide a solution, enabling 0-$\pi$ oscillations via Fermi energy tuning even in the absence of net magnetization~\cite{ouassou2023dc,lu2024varphi,beenakker2023phase,sun2025tunable,zhang2024finite}. Realizing controllable 0–$\pi$ transitions in antiferromagnetic JJs would thus pave the way for future superconducting electronics.

\begin{figure*}
		\centering
		\includegraphics[width=0.75\linewidth]{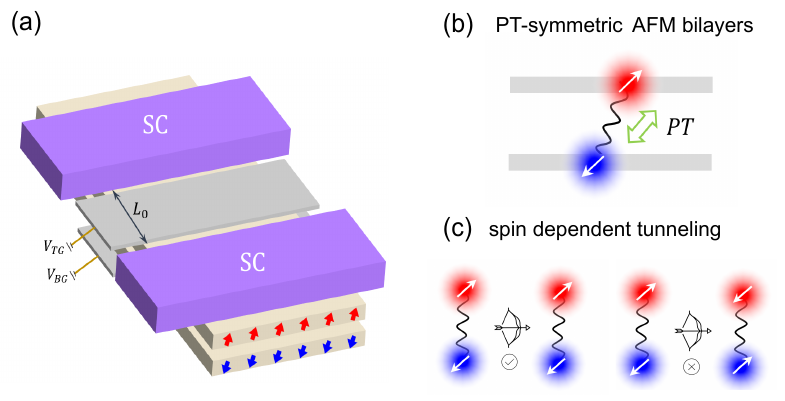}
		\caption{(a) A schematic picture of a lateral JJ based on $PT$-symmetric AFM bilayers. The AFM bilayers have the interlayer N\'{e}{e}l order. The two sides of the junction are superconducting (SC) electrodes. The weak-link region has dual gates $V_{TG}$ and $V_{BG}$ with length $L_0$. (b) In $PT$-symmetric AFM bilayers, the superconducting proximity effect causes the spin-layer-locked Cooper pairs. (c) Cooper pairs can tunnel between left and right domains when their internal N\'{e}el orders are aligned, but tunneling is forbidden when the N\'{e}el orders are opposite.}
		\label{fig:fig1}
\end{figure*}

$PT$-symmetric antiferromagnetic (AFM) bilayers, consisting of two coupled magnetic layers with antiparallel spin alignment, are of great interest for spintronics~\cite{sun2019giant,gao2021layer,fei2021pt,cao2023plane,wu2025spin,bobkov2025gate}. Recently, a pivotal experimental progress involves the electrostatic control of total magnetization in $PT$-symmetric AFM bilayer conductors, such as CrPS$_4$~\cite{yao2025switching}. It was observed in Ref.~\cite{yao2025switching} that a perpendicular displacement field can modulate spin polarization in these systems, where opposite spin polarizations are locked to the two layers because of $PT$ symmetry. An important question is whether 0–$\pi$ transitions can be realized in $PT$-symmetric AFM bilayer JJs, and how such junctions could enable new functional operations in superconducting electronics.

In this work, we first unveil a  mechanism of gate-controllable 0-$\pi$ oscillations in lateral JJs using $PT$-symmetric AFM bilayers [see Fig.~\ref{fig:fig1}(a)]. When proximitized to a conventional $s$-wave superconductor, the $PT$ symmetry of the system facilitates interlayer spin-singlet pairing via projection onto the low-energy subspace [see Fig.~\ref{fig:fig1}(b)]. Crucially, we show that the displacement field in the weak-link region induces an electrically tunable periodic 0-$\pi$ oscillations due to the unique interlayer Cooper pairing. More interestingly, because of the internal N\'{e}el order of the spin-layer-locked Cooper pairs, a Josephson giant magnetoresistor can be naturally designed: the supercurrent is strongly suppressed between antiphase interlayer cooper pairs. This occurs because Cooper pair tunneling becomes nearly prohibited when two sides of junctions exhibit antiparallel N\'{e}el ordering [see Fig.~\ref{fig:fig1}(c)]. Our theoretical analysis provides valuable insights for future studies of unconventional JJs based on $PT$-symmetric AFM bilayers.

\section{Microscopic Model}

To capture the essential low-energy physics of $PT$-symmetric AFM bilayers, we model the system with a minimal Hamiltonian in the fermionic basis $(\psi^b_{\uparrow},\psi^b_{\downarrow},\psi^t_{\uparrow},\psi^t_{\downarrow})^T$, which reads
\begin{equation}
\label{eq:eq_full_h}
H_0(\bm{k})=\lambda \bm{k}^2-\mu+g \tau_x \sigma_0 +J_{ex} \tau_z \bm{n}\cdot \bm{\sigma}.
\end{equation}
Here, $b$ ($t$) denotes the index of bottom (top) layer and $\lambda$ denotes the effective mass. $\mu$ is the Fermi energy and $g$ denotes the interlayer coupling. $J_{ex}$ represents the layer-contrasted exchange field, which characterizes the interlayer AFM order. The N\'{e}el order is $\bm{n}=(n_x,n_y,n_z)$ with $n_x=\sin\theta\cos\varphi, n_y=\sin\theta\sin\varphi, n_z=\cos\theta$. $\tau_i$ and $\sigma_i$ are the Pauli matrices in layer and spin space, respectively. 

\begin{figure*}
		\centering
		\includegraphics[width=1\linewidth]{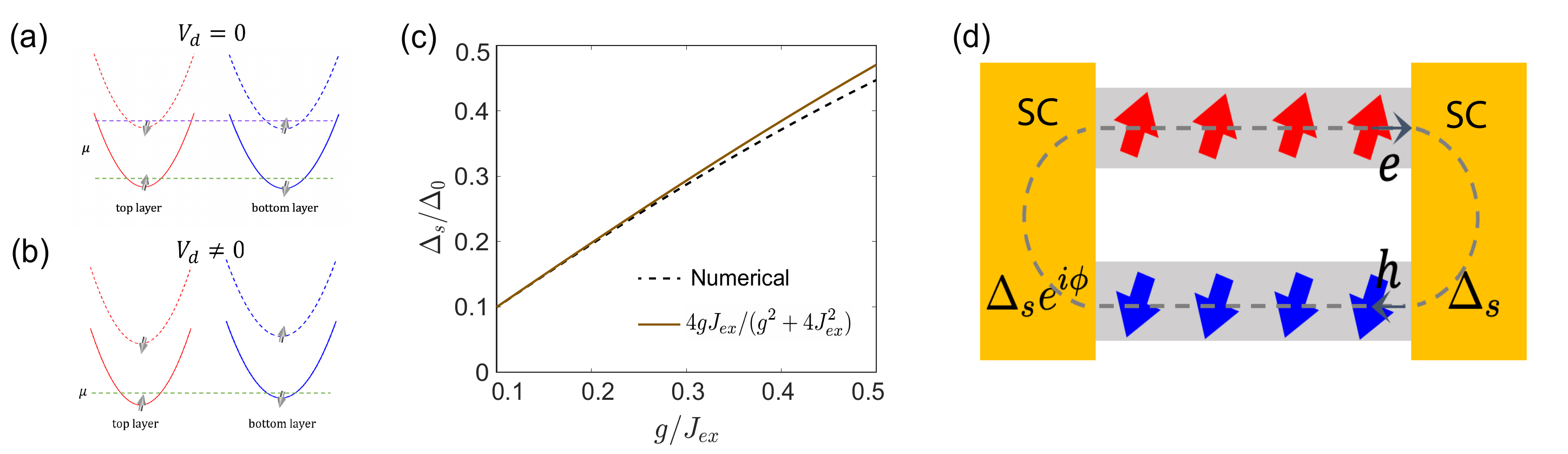}
		\caption{(a) In $PT$-symmetric AFM bilayers, an electron on the top layer has its $PT$-partner on the bottom layer, which are degenerate in energy. (b) When $V_d \neq 0$, the $PT$ symmetry is broken by lifting the band degeneracy. (c) The numerical and analytical results of the interlayer pairing gap. The numerical result is evaluated from Eq.~\eqref{eq:eq_tranfer}. (d) Schematic picture of the Andreev reflections of lateral JJ built by AFM bilayers. The Andreev reflection process involves an incoming spin-up electron from the top layer and an outgoing spin-down hole from the bottom layer.}
		\label{fig:fig2}
\end{figure*}

A schematic illustration of the designed lateral JJ with gate-defined weak-link is shown in Fig.~\ref{fig:fig1}, which has been experimentally realized~\cite{Rodan-Legrain2021, de2021gate,diez2023symmetry} and theoretically explored~\cite{Law2023, hu2023josephson,xie2023gate}. With the gate voltage $V_{TG}$ and $V_{BG}$, the additional Hamiltonian is $\delta H=V(x)\tau_z\sigma_0$ adding to Eq.~\eqref{eq:eq_full_h}. In this case, a layer potential difference $V(x)$ is induced by displacement field with $V(x)=V_d\Theta(x)\Theta(L_0-x)$. The $PT$ symmetry operator is given by $\hat{S}=\tau_x i\sigma_yK$ ($K$ is the complex conjugate), while $V(x)$ breaks $PT$ symmetry in the weak-link region and lift the band degeneracy within the weak-link [see Figs.~\ref{fig:fig2}(a) and (b)].  In Fig.~\ref{fig:fig1}(b) we can see that the Cooper pairs are contained by the $PT$ partners, yielding the interlayer Cooper pairing with the internal N\'{e}el order. 

We consider a lateral junction where $s$-wave superconductors are deposited on top of the AFM bilayers. The weak-link region is controlled by the potential $V(x)$. Due to the proximity effect, Cooper pairs tunnel into both the top and bottom layers; this scenario is valid under the assumption that the effective superconducting coherence length is much larger than the bilayer thickness ($d \lesssim 1$ nm). This process is modelled by adding the pairing potential $\hat{\Delta} = \Delta_0 \tau_0 i \sigma_y$ to $H_0$, where the spatial profile $\Delta(x) = \Delta_0 [ e^{i\phi} \Theta(-x) + \Theta(x - L_0) ]$ depends on the phase difference $\phi$ between the two superconductors.

Since the pairing is spin-singlet, electrons with opposite spins from different layers can form interlayer Cooper pairs. To describe this interlayer pairing explicitly, we derive an effective pairing potential near the Fermi energy. Note that $V_d=0$ at two sides of the junction so that the energy bands of the $PT$-symmetric AFM bilayers are doubly degenerate [Fig.~\ref{fig:fig2}(a)]. We consider the Fermi level to be near the band bottom, where interlayer pairing is prominent (indicated by the green labeling). In this regime, the low-energy effective Hamiltonian can be projected onto the basis $\psi=(\psi^t_{\uparrow},\psi^b_{\downarrow})^T$, with the chemical potential satisfying $\mu \gg \Delta_0$.

In $PT$-symmetric AFM bilayers, the wavefunctions for each layer without the interlayer coupling $g$ can be written as
\begin{align}
\psi_{t\uparrow} &= (0,0,e^{-i\frac{\varphi}{2}}\cos\frac{\theta}{2},e^{i\frac{\varphi}{2}}\sin\frac{\theta}{2})^T \\
\psi_{t\downarrow} &= (0,0,-e^{i\frac{\varphi}{2}}\sin\frac{\theta}{2} ,e^{i\frac{\varphi}{2}}\cos\frac{\theta}{2})^T
\end{align}
and 
\begin{align}
\psi_{b\uparrow} &= (e^{-i\frac{\varphi}{2}}\cos\frac{\theta}{2},e^{i\frac{\varphi}{2}}\sin\frac{\theta}{2},0,0)^T \\
\psi_{b\downarrow} &= (-e^{-i\frac{\varphi}{2}}\sin\frac{\theta}{2},e^{i\frac{\varphi}{2}}\cos\frac{\theta}{2},0,0)^T.
\end{align}
Here $b(t)$ denotes the bottom (top) layer. Focusing on the low-energy states localized on the top layer-spin up and bottom layer-spin down, thus in the presence of $g$ the wavefunctions are perturbed to
\begin{align}
u_1= \frac{1}{N_0}(\psi_{t\uparrow}-\frac{g}{2J_{ex}}\psi_{b\uparrow})\\
u_2= \frac{1}{N_0}(\psi_{b\downarrow}-\frac{g}{2J_{ex}}\psi_{t\downarrow}),
\end{align}
where $N_0=\sqrt{1+g^2/4J_{ex}^2}$ is the normalization factor. We can obtain the transformation matrix in the reduced Hilbert space $\tilde{U}(\theta,\varphi)=[u_1(\theta,\varphi),u_2(\theta,\varphi)]$.

The effective pairing in the reduced subspace can be obtained by 
\begin{equation}
\label{eq:eq_tranfer}
\tilde{\Delta}=\tilde{U}^\dagger(\theta,\varphi)\hat{\Delta}\tilde{U}^*(\theta,\varphi), 
\end{equation}
where $\tilde{U}(\theta,\varphi)=[u_1,u_2]$ is the projection operator and pairing matrix is $\hat{\Delta}=\Delta_0\tau_0 i\sigma_y$. $u_i$ are the eigenvectors of full Hamiltonian $H_0$. Treating the interlayer coupling $g$ as the perturbation term ($g \ll J_{ex}$), the resultant effective pairing can be derived as
\begin{equation}
\Delta_s=4g\Delta_0J_{ex}/(g^2+4J_{ex}^2).
\end{equation}
Due to spin-layer locking in this basis, $\Delta_s$ represents an effective interlayer pairing potential. As shown in Fig.~\ref{fig:fig2}(c), our analytical results for the interlayer pairing agree well with numerical calculations based on the full model in Eq.~\eqref{eq:eq_tranfer}. It is worth noting that when the Fermi level is high enough such that $\mu > 2\sqrt{J_{\text{ex}}^2 + g^2}$, the pairing involves not only interlayer Cooper pairs of the types ($b,\uparrow$)-($t,\downarrow$) and ($b,\downarrow$)-($t,\uparrow$) but also intralayer pairs. However, this is not the regime of interest for the present work. We therefore employ the identified interlayer pairing potential $\hat{\Delta}_{e}(x)= \Delta_s (i\sigma_y)[e^{i\phi}\Theta(-x)+\Theta(x-L_0)]$ to study the Josephson effect. In this description, the Pauli matrix $\sigma$ effectively acts on the layer pseudospin degree of freedom.

\begin{figure*}
		\centering
		\includegraphics[width=0.7\linewidth]{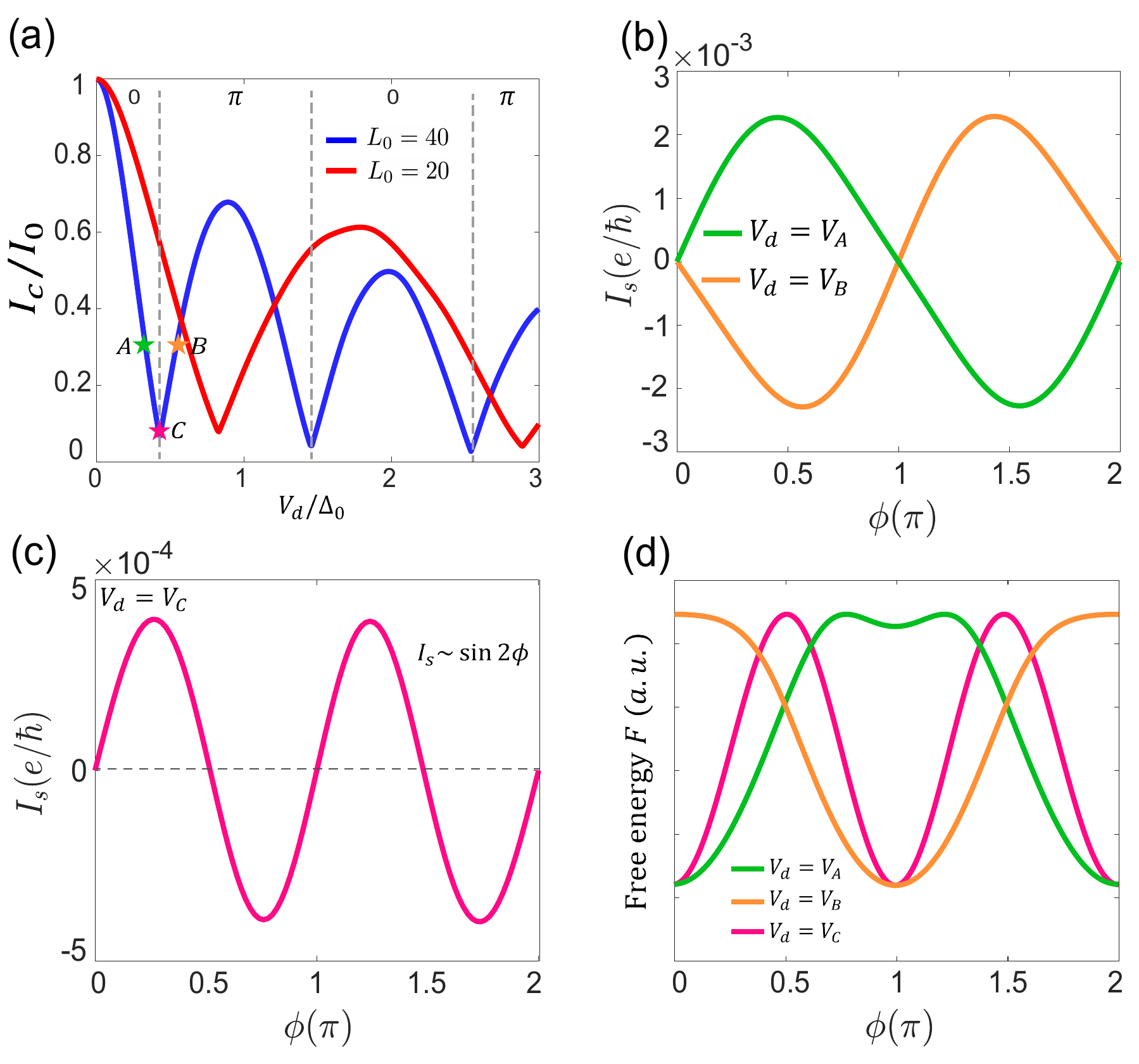}
		\caption{(a) $0-\pi$ oscillations: The maximum Josephson current $I_s$ as a function of the $V_d$ for $L_0=40,20$.  (b) The current-phase relation for $V_d=V_A$ and $V_B$, corresponding to the green and orange star in (a). (c) The $I_s$ near the $0-\pi$ transition (pink star in (a)). (d) The free energy for the three point in (d) with $0$, $\pi$, $0-\pi$ degenerate phases. Parameters for all panels: $\mu=0.5$, $\Delta_s=0.01$, $\Delta_0=0.05$ and $k_B T=0.3\Delta_s$.}
		\label{fig:fig3}
\end{figure*}

\section{Josephson current and 0-$\pi$ oscillations.}
The Josephson supercurrent is mediated by an Andreev reflection process involving an incoming spin-up electron from the top layer and an outgoing spin-down hole from the bottom layer. The full junction can be described by the Bogoliubov–de Gennes (BdG) Hamiltonian in the Nambu basis $(\psi,\psi^{\dagger})^T$ as
\begin{equation}
\label{eq:bdg_h}
\hat{H}(x)=\left(
\begin{matrix}{}
  &\hat{h}_{\alpha}(x)  & \hat{\Delta}_{e}(x)   \\
  &\hat{\Delta}^{\dagger}_{e}(x)  & - \hat{h}^*_{-\alpha}(x)
\end{matrix}\right),
\end{equation}
where $\hat{h}_{\alpha}(x)=-i v^{\alpha}_{f}\partial_x \sigma_0+ V(x)\sigma_z$, and $\sigma$ defined in the subspace $(\psi^t_{\uparrow},\psi^b_{\downarrow})^T$. Here $\alpha=\pm 1$ denote the right and left movers of the electrons. $v^{\alpha}_{f}=2\alpha \sqrt{\mu \lambda}\cos\theta_i$ is the fermi velocity and $\theta=\arctan(k_y/k_x)$ measures the injection angle. To be specific, we assume that the Fermi energy $\mu$ is identical between the superconductor and the weak link ($\mu_s=\mu_N=\mu$), resulting in $v_{f,s}=v_{f,w}=v_{f}$. In our calculations we adopt the temperature dependence of $\Delta_s$ with $\Delta_s(T)=\Delta_s \tanh(1.74\sqrt{T_c/T-1})$ and $\Delta_s=1.76k_B T_c$.

As shown in Fig.~\ref{fig:fig2}(d), the incoming wavefunction $\psi_{\text {in }}=[c_e^{-}(L), c_h^{+}(L), c_e^{+}(R),c_h^{-}(R)]^T$ and the outcoming wavefunction $\psi_{\text {out }}=[c_e^{+}(L), c_h^{-}(L), c_e^{-}(R),c_h^{+}(R)]^T$ satisfy the relations $\psi_{\text {out }}=\mathcal{S}_{\mathrm{A}}\psi_{\text {in }}$ and $\psi_{\text {in }}=\mathcal{S}_{\mathrm{N}}\psi_{\text {out }}$. The scattering matrices of the junction for the Andreev reflection $\mathcal{S}_A$ and the normal scattering $\mathcal{S}_N$ read  
\begin{equation}
\mathcal{S}_{\mathrm{A}}=\left(\begin{array}{cccc}
0 & e^{ i \frac{\phi}{2}-i \chi} & 0 & 0 \\
e^{- i \frac{\phi}{2}-i \chi} & 0 & 0 & 0 \\
0 & 0 & 0 & e^{- i \frac{\phi}{2}-i \chi} \\
0 & 0 & e^{ i \frac{\phi}{2}-i \chi} & 0
\end{array}\right),
\end{equation}
and 
\begin{equation}
\mathcal{S}_{\mathrm{N}}=\left(\begin{array}{cccc}
0 & 0 & e^{-ik_{e, l-} L_0} & 0 \\
0 & 0 & 0 & e^{-ik_{h, l+} L_0} \\
e^{ik_{e, l+} L_0} & 0 & 0 & 0 \\
0 & e^{ik_{h, l-} L_0} & 0 & 0
\end{array}\right)
\end{equation}
Here we take $L_0$ the junction length. If we only consider the linear order correction of Fermi momenta induced by the displacement field $V_d$, the correction can be written as $k_{e, l \alpha} \approx k_{N, l \alpha}^0+\delta k_{e, l \alpha}, k_{h, l \alpha} \approx  k_{N, l \alpha}^0+\delta k_{h, l \alpha}$ in which $k_{N, l \alpha}^0=\sqrt{\mu/\lambda}\cos \theta_i$ and 
\begin{equation}
\delta k_{e, l\alpha}=\frac{\epsilon-\delta\varepsilon_l}{v_{f}^{\alpha}}, \\\\ \delta k_{h, l\alpha}=\frac{\epsilon-\delta\varepsilon_l}{-v_{f}^{\alpha}}
\end{equation}
Here the energy shift $\delta\varepsilon_l=l(V_d+g_s\mu_B \bm{n}\cdot \bm{B})$ contains both the displacement field and magnetic field. Here $g_s$ is the Land\'{e} g-factor. In terms of $\hat{H}(x)$ in Eq.~\eqref{eq:bdg_h}, the supercurrent through the junction can be evaluated in the scattering matrix framework by~\cite{beenakker1991universal,brouwer1997anomalous,sun2024anomalous}
\begin{equation}
\label{eq:form_beenakar}
  I_s(\phi)=-\frac{2e}{\hbar\beta} \frac{d}{d \phi} \sum_{n=0}^{\infty} \ln \operatorname{det}\left[1-\mathcal{S}_A(i \omega_n) \mathcal{S}_N(i \omega_n)\right], 
  \end{equation} 
where $\omega_n=(2 n+1)\pi k_B T $ are fermionic Matsubara frequencies, $T$ is the temperature. Here we consider a wide sample $W\gg L_0$ with the junction width $W$. So there is the approximate translational invariance along $y$ axis and the preserved wave vector $k_y$. Then the Josephson current density can be evaluated by integrating the injection angle $\theta_{i}$, which reads
\begin{equation}
\label{eq:eq_int_i}
I_s(\phi)=-\frac{2e}{\hbar\beta} \int_{-\pi/2}^{\pi/2}d\theta_i\Gamma(\theta_i,\phi)
\end{equation}
with the integral function
\begin{equation}
\label{eq:eq_gamma_f}
\Gamma(\theta_i,\phi)=\sum_{n,l=\pm}\frac{\sin\phi}{\cos[2\chi-\frac{2(i\omega_n-lV_d)}{E(\theta_i)}]-\cos\phi},
\end{equation}
where $\chi=\arccos(i\omega_n/\Delta_s)$, $E(\theta_i)=2\sqrt{\mu\lambda}\cos(\theta_i)/L_0$. and $l=\pm$ denotes the summation over the layer index.  This general result contains the current from both the Andreev bound states and the continuum of states.

\begin{figure*}
		\centering
		\includegraphics[width=1\linewidth]{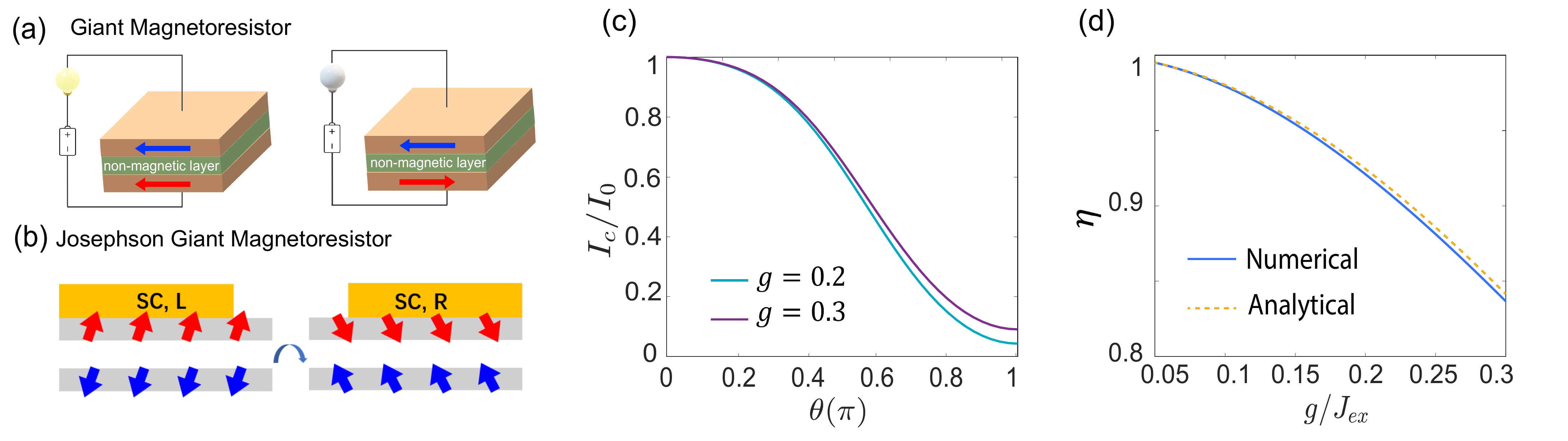}
		\caption{(a) The conventional giant magnetoresistor: for parallel  magnetization, the resistance is small with bright lamp; for antiparallel magnetization, the resistance is high with dark lamp. (b) The Josephson giant magnetoresistor: A lateral JJ with $PT$-symmetric AFM bilayers configurated as antiphase domains. (c) Critical current $I_c$ as a function of $\theta$ at interlayer coupling $g=0.2$ and $0.3$. (d) The JGMR quality factor $\eta$ versus $g$ with the numerical result (blue line) and the analytical result using Eq.~\eqref{eq:eq_eta} (yellow dashed line). We set the junction length to be $L_l=L_r=3$. Parameters: $\Delta_0=0.05$, $J_{ex}=1$, $k_B T=0.2\Delta_0$, $\mu=0.15$.}
		\label{fig:fig4}
\end{figure*}

Based on the formalism developed in Eqs.~\eqref{eq:eq_int_i} and \eqref{eq:eq_gamma_f}, we evaluate the Josephson supercurrent $I_s$ through the junction. Fig.~\ref{fig:fig3}(a) displays the critical current $I_c = \max(I_s)$ as a function of the displacement field $V_d$. As $V_d$ increases, $I_c$ exhibits damped oscillations with a $V_d$-dependent period, a hallmark signature of the $0$-$\pi$ transition.  Notably, the oscillation period approximately doubles when the junction length is reduced from $L_0 = 40$ to $L_0 = 20$. To further elucidate the nature of $0$-$\pi$ transition, Fig.~\ref{fig:fig3}(b) presents the phase dependence of $I_s$ at two highlighted displacement fields ($V_A = 0.3\Delta_0$ and $V_B = 0.6\Delta_0$), marked by stars $A$ and $B$ in Fig.~\ref{fig:fig3}(a). These points correspond to the $0$-junction and $\pi$-junction states, respectively. Crucially, the transition between these states can be achieved by finely tuning $V_d$ ($V_d \ll \lambda$), demonstrating purely electrical control over the phase.

Interestingly, near the transition [marked by the pink star in Fig.~\ref{fig:fig3}(a)], a pronounced second harmonic Josephson current emerges, as shown in Fig.~\ref{fig:fig3}(c), where the first harmonic term is strongly suppressed. The energy-phase relation $F(\phi)$ of the junction near the transition is given by $F(\phi) = E_1\cos\phi + E_2\cos 2\phi$, where $E_1$ and $E_2$ represent the first and second harmonic terms, respectively. Figure~\ref{fig:fig3}(d) displays the free energy $F$ for three characteristic points ($A$, $B$, and $C$), corresponding to ground states in the $0$, $\pi$, and degenerate $0$-$\pi$ regimes. Remarkably, in the regime where the $\cos 2\phi$ term dominates and the $0$ and $\pi$ states become energetically degenerate, the system can realize a $0$-$\pi$ qubit when the junction is shunted by a capacitor~\cite{blais2021circuit,koch2007charge,schreier2008suppressing}. This configuration offers the significant advantage of enabling a purely electronically controlled $0$-$\pi$ qubit, without requiring external magnetic field control~\cite{brooks2013protected,groszkowski2018coherence}.

\begin{figure*}
		\centering
		\includegraphics[width=0.9\linewidth]{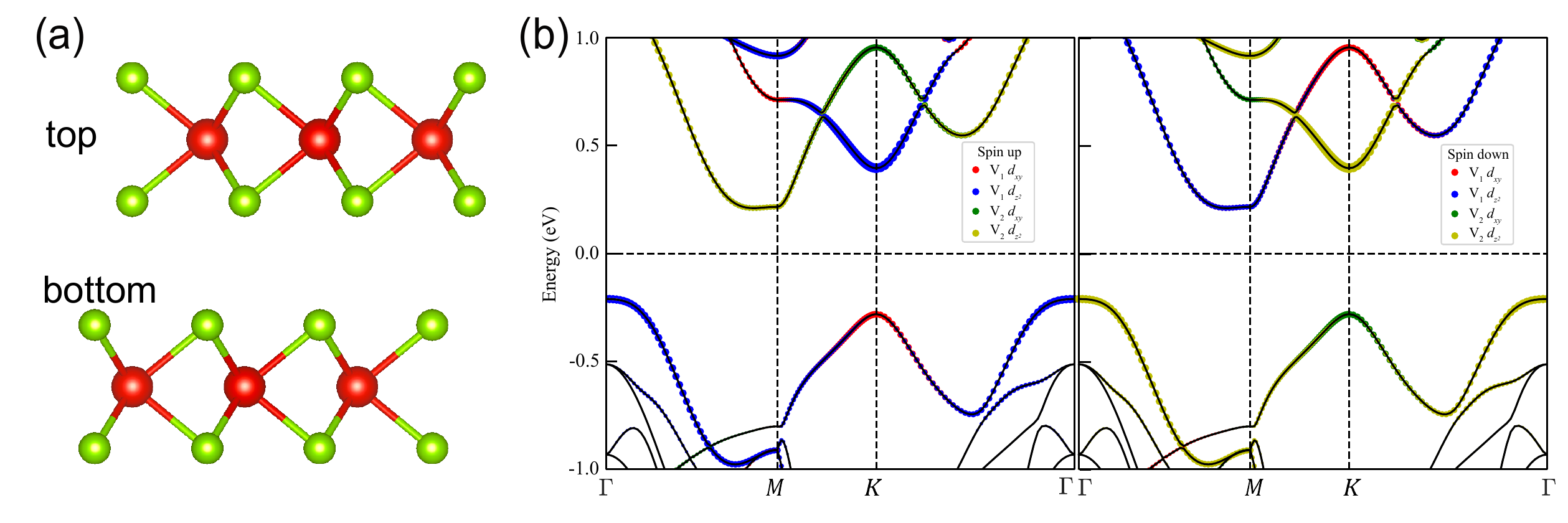}
    \caption{Crystal structure and band structure of VSe$_2$. (a) The crystal structure of bilayer VSe$_2$. The red and green spheres represent V and Se atoms, respectively. (b) The band structure of bilayer VSe$_2$ with projections on the $d_{xy}$ and $d_{z^2}$ orbitals of V atoms in different layers. The size of the markers represents the weight of the corresponding orbitals.}
		\label{fig:fig5}
\end{figure*}

We would like to point out that the $0$-$\pi$ oscillations studied here originate microscopically from field-induced finite-momentum pairing. In ferromagnetic junctions, spin splitting can generate Cooper pairs with finite center-of-mass momentum $Q =M/\sqrt{\mu \lambda}$ when a conventional superconductor is proximitized to a ferromagnetic weak link with $M$ the magnetization, leading to $\pi$-phase shifts with varying junction length or magnetization. In our system, despite the antiferromagnetic nature, the $PT$ symmetry enables dominate interlayer pairing. The displacement field induces a pseudo-Zeeman term in the layer space, which breaks the $PT$ symmetry and lifts the band degeneracy [see Fig.~\ref{fig:fig2}(b)].  Consequently, the spin-layer-locked Cooper pairs acquire finite momentum for small displacement field $V_d$
\begin{equation}
Q \approx V_d/\sqrt{\mu \lambda}.
\end{equation} 
The resulting gate-field-induced finite momentum drives the $0$-$\pi$ oscillations, analogous to the ferromagnetic case but with purely electrical control. 

Finally, we emphasize that the electronically controllable 0–$\pi$ junction transitions occur within an experimentally accessible regime. For example, for CrPS$_4$, the effective mass is $m^*=1.26m_e$~\cite{yao2025switching}, yielding $\lambda\approx 30$meV$\cdot$nm$^2$. Considering weak doping $\mu\approx 10$meV, the critical $V_d$ for the 0-$\pi$ transition can be estimated by $QL_0=\pi/2$, which gives $V_d\approx2$meV at $L_0=15$nm. This value is experimentally accessible~\cite{han2023orbital}.

\section{Josephson giant magnetoresistor}
We now point out another key property of these exotic spin-layer-locked Cooper pairs enabled by $PT$ symmetry: the Josephson giant magnetoresistance (JGMR) effect. In conventional giant magnetoresistor devices, the electrical resistance decreases when adjacent ferromagnetic layers have parallel magnetization alignments and increases significantly for antiparallel configurations [see Fig.~\ref{fig:fig4}(a)]. This spin-dependent transport arises from the relative orientation of the ferromagnetic layers' magnetization, which is the fundamental mechanism behind the tunneling magnetoresistance effect~\cite{baibich1988giant,camley1989theory,xiao1992giant}.

We propose here that such spin-dependent tunneling can similarly occur for Cooper pairs in JJs. As illustrated in Fig.~\ref{fig:fig4}(b), our designed junction incorporates configurable AFM domains---a configuration recently demonstrated experimentally in A-type bilayer AFM CrPS$_4$~\cite{wang2025configurable}, where the antiphase domains exhibit distinct N\'{e}el orders. Due to interlayer Cooper pairing, we predict that Cooper pairs can tunnel between left and right domains with:
\begin{itemize}
    \item Weak obstruction for parallel N\'{e}el order alignment
    \item Strong obstruction for antiparallel N\'{e}el order alignment
\end{itemize}
Remarkably, the supercurrent becomes strongly dependent on the internal spin texture despite the spin-singlet nature of the pairing [see Fig.~\ref{fig:fig1}(c)]. This enables a dissipationless JGMR effect, analogous to conventional GMR but occurring in the superconducting regime.

We first give a theoretical analysis of the JGMR effect. Assuming the tunneling matrix has the form $\hat{T}=\gamma \tau_0\sigma_0$, we can obtain the tunneling magnitude between Cooper pairs $|\mathcal{T}|^2$ by transforming the $\hat{T}$ in the subspace as $\hat{T}'=\tilde{U}^\dagger(\theta,\varphi)\hat{T}\tilde{U} (\theta,\varphi)$. We can obtain $|\mathcal{T}|^2$ by summing over all channels in $\hat{T}'$, yielding
\begin{equation}\label{eq:eq_tunnel}
|\mathcal{T}|^2=\frac{\gamma^2[1+g^2/J_{ex}^2+(1-g^2/J_{ex}^2)\cos\theta]}{2}.
\end{equation}
For the detailed derivations, see Supplementary Materials. Then, we can find the JGMR quality factor 
\begin{equation}
\label{eq:eq_eta}
   \eta=\frac{I_c(\theta=0)-I_c(\theta=\pi)}{I_c(\theta=0)+I_c(\theta=\pi)}= \frac{1-g^2/J^2_{ex}}{1+g^2/J^2_{ex}}.
\end{equation}
In the layered structure, we expect the interlayer coupling strength $g$ to be much smaller than the exchange field $J_{ex}$. As a result, the $\eta \approx 1 - 2g^2/J_{ex}^2$ can even approach one  with $ g \ll J_{ex}$. Physically, this indicates that a junction with parallel N\'eel order alignment is much more transparent than one with antiparallel N\'eel order, as expected. The factor $\eta$ decreases as $g$ increases, which arises from the fact that interlayer tunneling tends to weaken the spin-layer locking.

To verify this effect numerically, we calculate the Josephson supercurrent for various relative orientations $\theta$, where $\theta=0$ ($\theta=\pi$) corresponds to parallel (antiparallel) N\'{e}el order alignment. As shown in Fig.~\ref{fig:fig4}(c), the critical current $I_c$ exhibits significant suppression as $\theta$ approaches $\pi$, clearly revealing the nearly complete blockade of supercurrent in the antiparallel configuration. The corresponding quality factor $\eta\approx 90\%$, which matches our theoretical anticipation. Our results demonstrate that the JGMR effect can achieve a high quality factor. The unachievable $100\%$ quality factor stems from the finite interlayer coupling strength $g$. Notably, $I_c(\theta=\pi)$ exhibits a slight enhancement at larger $g$. We also plot Eq.~\eqref{eq:eq_eta} in Fig.~\ref{fig:fig4}(d) to compare the analytical and numerical results. At small $g$, Eq.~\eqref{eq:eq_eta} matches the numerical results well, but the two slightly deviate from each other as $g$ increases. This deviation arises from the perturbation approach used in the band projection method.

\section{Material candidate}
Based on the essential requirements for symmetry and spin-layer-locked features, we propose that the 2H phase with $AB$ stacking of the VSe$_2$ bilayer is a promising candidate material~\cite{gong2018electrically} for realizing the proposed Josephson junction. As shown in Fig.~\ref{fig:fig5}(a), two VSe$_2$ layers are stacked with intralayer ferromagnetic order and interlayer AFM order. The inversion symmetry, as a crystal symmetry, connects two V atoms in different layers, and the time-reversal symmetry flips the spin. Thus, the system preserves $PT$ symmetry.

We then analyze the $PT$ symmetry constraints on atomic orbital projections. For a single atomic orbital contribution, one can evaluate $\expval{\hat{O}_{\alpha}(\bm{r}-\bm{R_i})}_{n,\bm{k}} = \expval{P \hat{O}_{-\alpha}(\bm{r}-P\bm{R_i})}_{n,\bm{k}}$. This indicates that the eigenstates consist of $PT$-related orbitals with the same weight. With this in mind, we project the band structure onto the $d$-orbitals of V atoms in different layers. As shown in Fig.~\ref{fig:fig5}(b), all the bands are doubly degenerate over the entire Brillouin zone, and the states around the Fermi level are contributed by $d_{xy}$ and $d_{z^2}$ orbitals. Specifically, the spin-up $d_{z^2}$ orbitals from the top-layer V atoms (V$_1$) and the spin-down $d_{z^2}$ orbitals from the bottom-layer V atoms (V$_2$) dominate the states around the maximum valence band. Similarly, the conduction band minimum, localized at the $M$ point, is mainly contributed by the spin-up $d_{z^2}$ orbitals of V$_2$ and the spin-down $d_{z^2}$ orbitals of V$_1$. In addition to bilayer VSe$_2$, we predict other two-dimensional candidates, such as bilayer CrPS$_4$ and CrSBr, which are interlayer AFM and $PT$-symmetric. These materials exhibit similar band structure properties to those analyzed in bilayer VSe$_2$.

First-principles calculations are performed using the Vienna Ab initio Simulation Package (VASP) with the projector-augmented wave (PAW) method. The exchange-correlation functional is treated within the generalized gradient approximation (GGA) of Perdew-Burke-Ernzerhof (PBE). The energy cutoff for the plane-wave basis is set to 500 eV. The Brillouin zone was sampled using a 12 $\times$ 12 $\times$ 1 $\Gamma$-centered k-mesh. The GGA+U method is employed to account for the strong correlation effects of the 3d electrons of V atoms, with values of $U=2.0$ eV and $J=0.84$ eV. The lattice constants are taken from a previous study~\cite{gong2018electrically}, and the interlayer distance is optimized with a convergence criterion of 0.01 eV/\AA.

\section{Conclusion}

In summary, we have investigated unconventional Josephson effects in a superconductor/$PT$-symmetric AFM bilayers/superconductor JJ. The observed field-tunable $0$-$\pi$ oscillations emerge as a unique consequence of $PT$ symmetry and the layer structure. It is worth noting that in Eq.~\eqref{eq:eq_gamma_f} the 0-$\pi$ oscillations can also be contolled by the magnetic field (More details, see Supplementary Materials III~\cite{supp}). Furthermore, the unique N\'{e}el order of spin-singlet interlayer Cooper pairs enables a superconducting GMR effect, which can be controlled by manipulating Cooper pair tunneling between antiferromagnetic domains. Finally, we comment on candidate materials. A-type bilayer CrPS$_4$~\cite{yao2025switching}, VSe$_2$~\cite{gong2018electrically}, and CrSBr~\cite{wu2025spin} are promising platforms. In these systems, $P$ symmetry is broken but $PT$ symmetry is preserved, leading to the interlayer AFM order. Under a finite electric displacement field and with finite doping as realized in ref.~\cite{yao2025switching} recently, these AFM bilayers can become conductive via gating, thereby enabling the electrical control of 0-$\pi$ oscillations.

\section{Acknowledgments}
We thank helpful discussions with Ziting Sun, Song-Bo Zhang, Akito Daido, Yugui Yao and X. C. Xie. This work was supported by the Ministry of Science and Technology, China, and Hong Kong Research Grant Council through Grants No. 2020YFA0309600, No. RFS2021-6S03, No. C6025-19G, No. AoE/P-701/20, No. 16310520, No. 16307622, and No. 16309223. Y.M.X.  acknowledges financial support from the RIKEN Special Postdoctoral Researcher(SPDR) Program.

%

\end{document}